\documentstyle[12pt]{article}

\catcode`@=11
\def\dif{{\rm d}}
\def\deriv{\@ifnextchar[{\@deriv}{\@deriv[]}}
\def\@deriv[#1]#2#3{\mathchoice%
{{\dif^{#1}#2\over\dif{#3}^{#1}}}{{\dif^{#1}#2/\dif{#3}^{#1}}}%
{{\dif^{#1}#2\over\dif{#3}^{#1}}}{{\dif^{#1}#2/\dif{#3}^{#1}}}}
\def\derpar{\@ifnextchar[{\@derpar}{\@derpar[]}}
\def\@derpar[#1]#2#3{\mathchoice%
{{\partial^{#1}#2\over\partial{#3}^{#1}}}{{\partial^{#1}#2/\partial{#3}^{#1}
}}%
{{\partial^{#1}#2\over\partial{#3}^{#1}}}{{\partial^{#1}#2/\partial{#3}^{#1}
}}}

\def\secteqno{\@addtoreset{equation}{section}%
\def\theequation{\thesection.\arabic{equation}}}
\def\endsecteqno{\def\theequation{\@ifundefined{chapter}%
{\arabic{equation}}{\thechapter.\arabic{equation}}}}
\catcode`@=12
\secteqno
\def\beq{\begin{equation}}
\def\eeq{\end{equation}}
\font\sfit=cmssi10 scaled \magstep1

\newtheorem{theorem}{Theorem}
\newtheorem{corollary}{Corollary}
\newtheorem{proposition}{Proposition}
\newtheorem{definition}{Definition}
\newtheorem{lemma}{Lemma}

 1
\font\ddpp=msbm10 scaled \magstep 1
\def\R{\hbox{\ddpp R}} 

\let\ds=\displaystyle

\def\dif{{\rm d}}
\def\Dif{{\rm D}}
\def\Ker{\mathop{\rm Ker}\nolimits}

\def\Tan{{\rm T}}
\def\Ver{{\rm V}}
\def\LD{\hbox{\sfit L}}

\def\FD{{\cal D}}
\def\vl{{\rm vl}}
\def\invol{\hbox{\sfit s}}
\def\EL{{\cal E}}

\def\bfw{{\bf w}}
\let\eps=\varepsilon

\begin{document}

\vskip 3mm

\begin{center}
{\large \bf
SOME GEOMETRIC ASPECTS OF VARIATIONAL CALCULUS IN CONSTRAINED SYSTEMS}
\\[7mm]
Xavier Gr\`{a}cia\\[0.5mm]
{\it Departament de Matem\`{a}tica Aplicada i Telem\`{a}tica\\
Universitat Polit\`{e}cnica de Catalunya\\
Campus Nord UPC, edifici C3, C. Jordi Girona 1, 08034 Barcelona,
Spain\\[0.5mm]
{\tt xgracia@mat.upc.es}}\\[4mm]
Jes\'us Mar{\'\i}n-Solano\\[0.5mm]
{\it Departament de Matem\`atica Econ\`omica, Financera i Actuarial\\
Universitat de Barcelona\\
Av. Diagonal 690, 08034 Barcelona, Spain\\[0.5mm]
{\tt jmarin@eco.ub.es}}\\[4mm]
Miguel-C. Mu\~{n}oz-Lecanda\\[0.5mm]
{\it Departament de Matem\`atica Aplicada i Telem\`{a}tica\\
Universitat Polit\`{e}cnica de Catalunya\\
Campus Nord UPC, edifici C3, C. Jordi Girona 1, 08034 Barcelona,
Spain\\[0.5mm]
{\tt matmcml@mat.upc.es}}
\vskip 4mm
\end{center}

\pagestyle{myheadings}
\markright{X.\,Gr\`acia \it et al
\quad
Variational calculus in constrained systems}

\begin{abstract}
We give a geometric description of variational principles in mechanics,
with special attention to constrained systems. For the general case of
nonholonomic constraints, a unified variational approach is given, and the
equations of motion of both vakonomic and nonholonomic frameworks are
obtained. We study specifically the existence of infinitesimal variations in
both cases.
When the constraints are integrable,
both formalisms are compared and it is proved that they coincide.
As examples, we give geometric formulations of the equations of motion
for the case of optimal control and
for vakonomic and nonholonomic mechanics
with constraints linear in the velocities.
\end{abstract}


\newpage

\section{Introduction}

To describe the motion of mechanical systems there is a variety of
mathematical models which are based on different principles.
Most of the physical models are obtained using an appropriate variational
principle in a certain evolution space.
But variational principles are not only important in physics, but also in
many branches of engineering \cite{Ze} or economics \cite{In}, where one is
interested in optimizing a given functional, possibly subject to some
restrictions.
In fact, constraints are ubiquitous in many mechanical systems and much more
different situations.

\medskip

In this paper we are going to study Lagrangian systems, i.e., dynamical
systems in which the equations of motion are obtained by finding the
critical paths of a functional
\[
\int_{t_1}^{t_2} L \,\dif t\; ,
\]
where $L$ is a function defined on the tangent bundle $\Tan Q$ of a given
differentiable manifold $Q$, the configuration space. We will not consider
arbitrary variations, but only variations satisfying certain conditions.
These conditions arise from some given constraints on the dynamics of the
system. We will analize the case when the constraints are defined by a
certain submanifold $C$ of the tangent bundle $\Tan Q$. Such constraints are
usually called nonholonomic.

\medskip

There are two different approaches when dealing with constraints. The first
one is based on the idea of understanding the constraints as constraint
forces. This point of view, that seems very natural in a physical context,
gives rise to the classical d'Alembert-Lagrange principle. Mechanics of
Lagrangian systems with nonholonomic constraints based on this principle is
called nonholonomic mechanics \cite{Ar}\cite{Ga}\cite{Ve}. But there is
another different point of view, that seems more natural when one is
interested in optimizing a given functional defined as above when there are
constraints. For example, if we wish to change the state of a given system
minimizing a cost functional (a typical problem in engineering or
economics), it is not natural to understand the constraints as forces acting
on the system. In this case one is interested in minimizing the functional
considering only the variations allowed by the constraints. Mechanics of
Lagrangian systems with nonholonomic constraints based on this idea is often
called vakonomic mechanics (mechanics of {\bf v}ariational {\bf a}xiomatic
{\bf k}ind \cite{Ar}). For example, optimal control theory is a typical
example of vakonomic mechanics. It is interesting to notice that both
mechanics do not coincide in general, but they agree when the constraints
are integrable. Several references on these topics are \cite{Ar},
\cite{Bli}, \cite{Els} and \cite{Le}.
See also
\cite{LMM,SC}.

\medskip

The paper is organized as follows. In section 2 we give a generalized notion
of a variational problem. Section 3 is devoted to study vakonomic mechanics
from a geometric point of view, with special attention to the existence of
admissible variations in order to obtain the equation of motion. In section
4 we do the same for nonholonomic mechanics. Both mechanics are understood
as generalized variational problems. In section 5 it is proved that, when
the constraints are integrable, vakonomic and nonholonomic mechanics
coincide. In section 6, as an example of vakonomic mechanics, a geometric
formulation of optimal control theory is studied. Finally, in section 7, we
give a geometric formulation of vakonomic mechanics when the constraints are
defined by a distribution (linear constraints in the velocities).

\medskip

Basic knowledge of differential geometric structures is assumed. The
presentation is almost self-contained but the interested reader may consult
the bi\-blio\-gra\-phy for more specific topics as the vertical lift, the
fibre derivative and the canonical involution \cite{Ab}, or the
Euler-Lagrange operator \cite{Ca}.

\section{Variational problems}

\subsection{Elements of a variational problem}

First, we are going to define what we mean by a variational problem. A
variational problem consists of the data $(Q,L,C,{\cal C},{\cal W})$ where:
\begin{itemize}
\item $Q$ is a $n$-dimensional differentiable manifold, the {\it
configuration space}.

\item $L$ is the {\it lagrangian function} defined on the tangent bundle,
$L \colon \Tan Q\longrightarrow\R$.

\item $C$ is the {\it constraint submanifold}, and it is a submanifold of
$\Tan Q$.

\item $\cal{C}$ is the family of {\it admissible paths}. Given two points
$q_1, q_2 \in Q$
and a compact interval $I = [t_1,t_2]$, we will say that a path $\gamma
\colon I \to Q$ of class $C^2$ is admissible if:

$\gamma(t_1) = q_1$, $\gamma(t_2) = q_2$ and

$\dot{\gamma}(t)\in C$, for all $t\in I$.

\item ${\cal W}$ are the {\it admissible variation fields} (or infinitesimal
variations). For a given admissible path $\gamma$, ${\cal W}_\gamma$
consists on a certain set of $C^1$ vector fields along~$\gamma$.
\end{itemize}

Notice that we do not consider variations of a path $\gamma$, but variation
fields along~$\gamma$.
\smallskip
Now we are ready to define the variational problem associated to
$(Q,L,C,\cal{C},\cal{W})$. The {\it action}\/ of $L$ along a path $\gamma$
is the functional $S \colon {\cal C}\longrightarrow\R$ given by the integral
\beq\label{accion}
S[\gamma] = \int_I L(\dot\gamma(t)) \,\dif t\; .
\eeq

A variational problem consists in finding the critical admissible paths of
the functional $S$, in a sense that will be precised later.

\subsection{Variations and variation fields}

Let $\gamma \colon I\longrightarrow Q$, $\gamma(t_1)=q_1$,
$\gamma(t_2)=q_2$, be an admissible path.
A {\it variation}\/ of $\gamma$ is a ${\rm C}^2$ function
$\Gamma \colon (-\delta,\delta)\times I\longrightarrow Q$ such that:
\begin{enumerate}
\item $\Gamma_\eps = \Gamma(\eps,\cdot)$ is a one-parameter family of paths
defined on $I$
with fixed end-points, $\Gamma(\eps,t_i)=q_i$,
$\forall\eps\in(-\delta,\delta)$, $i=1,2$, and
\item $\Gamma(0,t) = \gamma(t)$, $\forall t\in I$ (if there is no variation,
$\eps=0$, we obtain the original path $\gamma$).
\end{enumerate}

Given a function $\Gamma(\eps,t)$ of two real variables, we will denote its
derivatives with respect to $\eps$ and $t$ as $\Gamma'$ and $\dot\Gamma$,
respectively. It is clear that $\Gamma'$ and $\dot\Gamma$ are vector fields
along~$\Gamma$.

Now, we are in conditions to define a variation field of $\Gamma$.

\begin{definition} Given a variation $\Gamma(\eps,t)$ of an admissible path
$\gamma$, the {\em variation field} of $\Gamma$ is the vector field ${\bf
w}$ along $\gamma$ defined by
\[
\bfw(t)= \Tan_{(0,t)}\Gamma \cdot \left. \deriv{}{\eps} \right|_{(0,t)} =
\Gamma'(0,t)\; .
\]
\end{definition}

Notice that
$$
\bfw(t_1) = 0 , \quad \bfw(t_2) = 0\; ,
$$
since the $\Gamma_\eps$ have fixed end-points.

Therefore, given a family of variations of $\gamma$, we can associate to
them a family of variation vector fields along~$\gamma$. We will say that a
variation $\Gamma$ of $\gamma$ is {\it admissible} if its associated
variation vector field along~$\gamma$, $\bfw(t)=\Gamma'(0,t)$, is an
admissible variation field of $\gamma$, i.e., $\bfw\in{\cal W}_\gamma$.

\medskip

We finish this description about variations and variation fields with a
useful lemma, whose proof is straighforward in local coordinates.

\begin{lemma}\label{lambda-w} For any $\bfw$ vector field along $\gamma$ and
any function $\lambda\colon I\to\R$,
\[
\invol \circ (\lambda \bfw)^{\textstyle.} =
(\Dif \lambda)\, \vl(\dot\gamma,\bfw) + \lambda \,\invol \circ \dot\bfw\; .
\]
\hfill $\Box$
\end{lemma}

Notice that, if $\bfw$ is a vector field along~$\gamma$, then $\dot\bfw$ is
a vector field along $\bfw$ and $\invol \circ \dot\bfw$ is a vector field
along~$\dot\gamma$. The function $\lambda \colon I \to \R$ denotes a
function of time, and it is clear that $\lambda \bfw$ is another vector
field along~$\gamma$. $\Dif$ is the usual derivative with respect to the
time. The map $\vl$ denotes the vertical lift $\vl\colon\Tan Q\times_Q\Tan
Q\to\Tan(\Tan Q)$. Its local expression is $\vl (q,v,u)=(q,v;0,u)$. Finally,
$\invol \colon \Tan(\Tan Q) \to \Tan(\Tan Q)$ denotes the canonical
involution, which is an isomorphism between the two vector bundle structures
of $\Tan(\Tan Q)$. Its local expression is $\invol(q,v;u,a) = (q,u;v,a)$.

\subsection{Critical admissible paths}

\begin{definition} An admissible path $\gamma$ is said to be {\em critical}
if, for each admissible variation $\Gamma_\eps$, the first variation of
$S[\Gamma_\eps]$ is zero; i.e.,
\[
\left. \deriv{}{\eps} S[\Gamma_\eps] \right|_{\eps=0} = 0
\]
for each admissible variation $\Gamma_\eps$ of $\gamma$.
\end{definition}

The main purpose of this paper consists in discussing the criticity
conditions for different variational problems and describing the solutions.
First of all, we are going to describe the criticity condition for a general
problem.

It is clear that, if $g \colon Q \to \R$ is a function,
then, for any function $\Gamma(\eps,t)$ ($\Gamma \colon
U\subset\R^2\longrightarrow Q$) of two real variables,
$$
\derpar{}{\eps} g(\Gamma(\eps,t)) =
\langle \dif g(\Gamma(\eps,t)) , \Gamma'(\eps,t) \rangle\; ,
$$
and similarly for $\derpar{}{t}$.

\medskip

Now, suppose that $\Gamma(\eps,t)$ is a variation of a path $\gamma$. Let us
consider $\dot\Gamma \colon (-\delta,\delta) \times I \to \Tan Q$.
Derivation of $\dot\Gamma$ with respect to~$\eps$ and~$t$ yields
$(\dot\Gamma)'$ and $(\dot\Gamma)^{\textstyle.}$, which are now vector
fields along~$\dot\Gamma$. Taking $\eps=0$ yields two vector fields
along~$\dot\gamma$, which are $\invol \circ \dot{\bfw}$ and $\ddot\gamma$.

Then, if $f \colon \Tan Q \to \R$ is a function, we have
\beq
\left. \derpar{}{\eps} \right|_{\eps=0} f(\dot\Gamma(\eps,t)) =
\langle \dif f(\dot\gamma(t)) , \invol(\dot\bfw(t)) \rangle\; .
\label{epsder}
\eeq
(Remember that $\dot\bfw$ is a vector field along~$\bfw$,
and $\invol \circ \dot\bfw$ is a vector field along~$\dot\gamma$,
so the contraction makes sense.)

Now, we can characterize the criticity condition in a more manageable way.

\begin{proposition}\label{PEL} Given a variational problem
$(Q,L,C,\cal{C},\cal{W})$, an admissible path $\gamma$ is critical if and
only if
\[
\int_I
\left\langle \dif L(\dot\gamma(t)) , \invol(\dot\bfw(t)) \right\rangle
\,\dif t\ = 0\; ,
\]
for each admissible vector field $\bfw\in{\cal W}_\gamma$.
\end{proposition}

{\bf Proof:} Using (\ref{epsder}) in (\ref{accion}), we obtain
\[
\left. \deriv{}{\eps} S[\Gamma_\eps] \right|_{\eps=0}
=
\int_I
\left\langle \dif L(\dot\gamma(t)) , \invol(\dot\bfw(t)) \right\rangle
\,\dif t\; ,
\]
and the result follows.
\hfill $\Box$

\bigskip

Observe that this condition does not depend on the full variation
$\Gamma(\eps ,t)$, but only on its variation field (see also \cite{Calo}).
Therefore, in our study of variational calculus, we will shift our attention
to infinitesimal variations rather to finite variations.

\subsection{The Euler-Lagrange operator}

To obtain a more manageable condition of criticity, it is convenient to
perform an integration by parts. First, let us define the Euler-Lagrange
operator of $L$.

\begin{definition}\label{EL} The {\em Euler-Lagrange operator} associated
with a function $L \colon \Tan Q\longrightarrow\R$ is a mapping $\EL_L
\colon \Tan^2Q \to \Tan^*Q$ defined by the relation
\[
\langle \EL_L \circ \ddot\gamma , \bfw \rangle
=
\langle \dif L \circ \dot\gamma , \invol \circ \dot\bfw \rangle
-
\Dif \langle \FD L \circ \dot\gamma , \bfw \rangle\; ,
\]
for any path $\gamma\colon I\to Q$ and vector field $\bfw$ along~$\gamma$.
\end{definition}

Here, the map $\FD L \colon \Tan Q \to \Tan^*Q$ is the fibre derivative of
$L$. Recall that, given a vector bundle $E \to B$, if $f \colon E \to \R$ is
a function, then the derivatives of the restrictions of $f$ to the fibres
define the fibre derivative of~$f$, which is a map $\FD f \colon E \to E^*$.
Its local expression is $\FD f(b;a) = (b; \derpar fa)$.

It is easy to check (in coordinates) that the Euler-Lagrange operator is
well-defined by this relation.

Therefore, the Euler-Lagrange operator is a one-form along the projection
$\Tan^2Q \to Q$, and also an affine bundle map along $\Tan Q \to Q$. The
expression in local coordinates of $\EL_L$ is the usual one,
$$
\EL_L =
\left( \derpar Lq - \deriv{}{t}\left( \derpar Lv \right) \right) \dif q\; ,
$$
where $\deriv{}{t}$ is the total time-derivative operator. The
Euler-Lagrange operator can be extended in the same way to a time dependent
Lagrangian.

\medskip

Using the definition of the Euler-Lagrange operator, we can characterize the
criticity condition in the usual form.

\begin{theorem}\label{ELgen} Given a variational problem
$(Q,L,C,\cal{C},\cal{W})$, an admissible path $\gamma$ is critical if and
only if
\[
\int_I \left\langle \EL_L(\ddot\gamma(t)) , \bfw(t) \right\rangle \,\dif t
=0\; ,
\]
for each admissible variation field $\bfw\in{\cal W}_\gamma$.
\end{theorem}

{\bf Proof:} From proposition \ref{PEL} and the definition \ref{EL} of the
Euler-Lagrange ope\-ra\-tor we obtain that
$$
\left. \deriv{}{\eps} S[\Gamma_\eps] \right|_{\eps=0}
=
\int_I \left\langle \EL_L(\ddot\gamma(t)), \bfw(t) \right\rangle \,\dif t +
\bigg[\langle \FD L(\dot\gamma(t)) , \bfw(t) \rangle \bigg]_{t_1}^{t_2}\; .
$$
The result follows observing that, since $\bfw$ is a variation field, the
last term vanishes ($\bfw(t_1)=0$, $\bfw(t_2)=0$).
\hfill $\Box$

\bigskip

For the case when there are no constraints, $C=\Tan Q$ and ${\cal W}_\gamma$
is the set of all the vector fields along~$\gamma$, we obtain the well-known
{\it Euler-Lagrange equation}.

\begin{corollary} Given a unconstrained variational problem, a path
$\gamma\in{\cal C}$ is critical if and only if
\[
\EL_L \circ \ddot\gamma = 0\; .
\]

\end{corollary}

In this paper we will be interested in variational problems when $C\subseteq
\Tan Q$, $C\neq \Tan Q$. Hence, given a set of admissible paths, it is
necessary to select a set of admissible variation fields (or infinitesimal
variations) along the admissible paths. We will consider two different
approaches to this problem. The first one is vakononomic mechanics, which
can be considered as a strictly variational approach. The second one is
nonholonomic mechanics, which is variational in our generalized sense, but
not in the classical one. Nonholonomic mechanics is the usual way to
describe the dynamics of a mechanical system with constraints. In the next
two sections we will describe the dynamical equations obtained in each case.
It is interesting to remark that both approaches are equivalent when the
constraints are integrable (holonomic constraints).

\medskip

We finish this section with a useful property of the Euler-Lagrange operator
that will be used in many calculations in the following. The proof is
straighforward in local coordinates.

\begin{lemma}\label{mu-f}
For any $f \colon \Tan Q \to \R$ and $\mu \colon I \to \R$ (a function of
time),
\[
\EL_{\mu f} = \mu\, \EL_f - (\Dif \mu) \FD f \circ \tau^2_{\,1}\; ,
\]
where $\tau^2_{\,1} \colon \Tan^2 Q \to \Tan Q$ is the canonical projection.
\hfill $\Box$
\end{lemma}

\section{Vakonomic mechanics}

Roughly speaking, vakonomic mechanics is the result of variational calculus
when the variations are restricted by some constraints on the positions and
also the velocities.

Our initial setting is therefore a submanifold $C \subset \Tan Q$ of
codimension~$m < n$;
let us denote by $j$ the inclusion of $C$ in $\Tan Q$. A {\it constraint} is
any function $\phi$ vanishing on $C$. Locally $C$ is defined by the
vanishing of some constraints $\phi^i \colon \Tan Q \to \R$ ($i=1,\dots ,m$)
whose differentials $\dif \phi^i$ are linearly independent at each point of
$C$.

We will assume that {\it the projection of $C$ to~$Q$,
$\tau_Q \circ j \colon C \longrightarrow Q$, is a submersion}.
It can be easily proved that this statement is equivalent to say that the
constraints $\phi^i$ can be chosen such that their fibre derivatives $\FD
\phi^i$ are linearly independent at every point of $C$. In local
coordinates, this means that $\derpar{\phi^i}{v^k}$ has maximal rank. That
is, the constraints restrict the velocities, not the positions.

With the assumptions above, the image $(\tau_Q\circ j)(C) \subset Q$ is
open, so we may assume that the projection $C \to Q$ is a {\it surjective}\/
submersion. Then there exists the vertical subbundle $\Ver(C) \subset
\Tan(C)$, which has rank $n-m$ (the dimension of the fibres of the
submersion). Indeed, at each $v_q \in C$ we have $\Ver_{v_q}(C) =
\Tan_{v_q}(C) \cap \Ver_{v_q}(\Tan Q)$.

To obtain the equations of motion of vakonomic mechanics, we need first to
describe which are the admissible variations.

\subsection{The variations of vakonomic mechanics}

We remember that an admissible path is a mapping $\gamma \colon I \to Q$
such that $\dot\gamma$ takes its values in the submanifold $C \subset \Tan
Q$. Due to our assumptions on $C$, there exist vector fields locally defined
on $Q$ taking values in $C$. Their integral curves have their derivatives in

$C$, so there are many admissible paths.

\begin{definition}
Let $\Gamma$ be a variation of an admissible path~$\gamma$. The variation
$\Gamma$ is called a {\em strongly admissible variation} of $\gamma$ if
every path $\Gamma_\eps$ is admissible.
\end{definition}

If $\Gamma$ is a strongly admissible variation, then
$\phi(\dot\Gamma(\eps,t)) = 0$, for any constraint~$\phi$. Taking the
derivative with respect to $\eps$ at $\eps=0$ and using (\ref{epsder}) we
have
\[
\langle \dif \phi \circ \dot\gamma , \invol \circ \dot\bfw \rangle = 0
\]
for every constraint~$\phi$.
This can also be expressed as
\[
\invol(\dot\bfw(t)) \in \Tan_{\dot\gamma(t)}(C)
\]
for each $t \in I$.

\begin{definition}\label{varvak}
A variation field $\bfw$ of an admissible path $\gamma$
is called an {\em admissible variation field} for a given vakonomic problem
if
\[
\invol(\dot\bfw(t)) \in \Tan_{\dot\gamma(t)}(C)\; ,
\]
that is, $\langle \dif \phi \circ \dot\gamma , \invol \circ \dot\bfw \rangle
= 0$.
\end{definition}

From the definition of the Euler-Lagrage operator (definition \ref{EL}), we
obtain that $\bfw$ is a variation field if and only if
\[
\langle \EL_\phi \circ \ddot\gamma , \bfw \rangle
=
- \Dif \langle \FD \phi \circ \dot\gamma , \bfw \rangle\
\]
for every constraint $\phi$.

It is important to remark that an admissible path may not have any
nontrivial strongly admissible variation, and so an admissible variation
field may not arise from a strongly admissible variation. One may say that
the variations having admissible variation fields are the variations that
preserve the constraints up to first order in~$\eps$. These variations may
be called {\it weakly admissible variations}.

\bigskip

Next, we are going to give a more detailed description of admissible
variation fields. Among all the vector fields $\bfw$ along~$\gamma$, we
consider a particular submodule. Take the subbundle
$$
\LD_\gamma^C \subset \gamma^* \Tan(Q)=I\times_\gamma \Tan (Q)
$$
whose sections are the vector fields $\bfw$ along~$\gamma$ of class $C^1$
whose vertical lifts $\vl(\dot\gamma,\bfw)$ are tangent to $C$. Using this
subbundle we can express the admissible variation fields in a more
manageable way. First, notice that, since $I$ is an interval, both
$\LD_\gamma^C$ and $\gamma^* \Tan (Q)$ are trivializable. Therefore there
exists a global frame for $\gamma^* \Tan (Q)$, $(\bfw_k)$ ($k=1,\dots ,n$).
Since $\LD_\gamma^C$ is a subbundle of rank $n-m$, we can assume that the
last $n-m$ of the $\bfw_k$
span this subbundle.

Any vector field along $\gamma$ can be thus uniquely written $\bfw =
\sum_{k=1}^{n} \lambda^k \bfw_k$, where $\lambda^k$ are functions of time.
Then $\bfw$ is a variation field if and only if the coefficients $\lambda^k$
vanish at the end-points of $I$. Moreover, according to definition
\ref{varvak}, $\bfw$ is an admissible variation field if it is a variation
field and $\langle \dif \phi^i \circ \dot\gamma, \invol \circ \dot\bfw
\rangle = 0$, for $i=1,\dots ,m$. Taking into account lemma \ref{lambda-w},
this condition can be written
\beq\label{des}
\sum_{k=1}^{n}
\langle \FD \phi^i \circ \dot\gamma, \bfw_k \rangle \Dif \lambda^k
+
\sum_{k=1}^{n}
\langle \dif \phi^i \circ \dot\gamma, \invol \circ \dot\bfw_k \rangle
\lambda^k
= 0\; .
\eeq
Notice that, since the fibre derivatives $\FD \phi^i$ are linearly
independent at each point,
the matrix with entries $\langle \FD \phi^i \circ \dot\gamma, \bfw_k
\rangle$ has maximal rank,~$m$. By the special choice of the $\bfw_k$, the
last $n-m$ of them vanish under the $\FD \phi^i$, and hence the square
matrix $A = (\langle \FD \phi^i \circ \dot\gamma, \bfw_j
\rangle)_{i,j=1,\dots ,m}$ is invertible.
So, writing the equation as
\begin{equation}\label{humm}
\sum_{j=1}^{m} A^i_{\,j} \, \Dif \lambda^j +
\sum_{j=1}^{m} B^i_{\,j} \, \lambda^j + \sum_{l=m+1}^{n} C^i_{\,l} \,
\lambda^l
= 0\; ,
\end{equation}
we can isolate the $\Dif \lambda^j$ ($j=1,\dots ,m$) linearly in terms of
all the $\lambda^k$ ($k=1,\dots, n$). This determines uniquely $\lambda^j$,
$j=1,\dots,m$ as functions of $\lambda^l$, $l=m+1,\dots,n$, due to the
initial condition $\lambda^j(t_1)=0$. However, not any $\lambda^l$,
$l=m+1,\dots,n$ vanishing on $t_1,t_2$ are admissible. Notice that the
solutions $\lambda^j$, $j=1,\dots, m$ must vanish also in $t_2$. In fact,
the existence of solutions of (\ref{humm}) satisfying
$\lambda^k(t_1)=\lambda^k(t_2)=0$, $k=1,\dots n$, is not guaranteed in
principle. If we write (\ref{humm}) as
\[
A \, \dot\lambda_{(1)} = -
B \, \lambda_{(1)} -
C \, \lambda_{(2)}\; ,
\]
the solution satisfying $\lambda_{(1)}(t_1)=0$ is
\[
\lambda_{(1)}(t)=-\nu(t)\int^t_{t_1} [\nu(s)]^{-1}\, A^{-1}(s)\,
C(s)\,\lambda_{(2)}(s)\,\dif s\; ,
\]
where $\nu(t)$ is the fundamental matrix of the homogeneous system $A \,
\dot\lambda_{(1)} = -
B \, \lambda_{(1)}$ satisfying the initial condition
$\nu^j_{\,i}(t_2)=\delta^j_{\,i}$. If $\lambda_{(1)}(t_2)=0$, then
necessarily
\begin{equation}\label{uultima}
\int^{t_2}_{t_1} [\nu(s)]^{-1}\, A^{-1}(s)\, C(s)\,\lambda_{(2)}(s)\,\dif s
= 0\; .
\end{equation}
If the system is homogeneous ($C=0$) we obtain the trivial solution
$\lambda^j(t)=0$, $j=1,\dots,m$, remaining $\lambda^l$, $l=m+1,\dots,n$ as
arbitrary functions satisfying the boundary conditions
$\lambda^l(t_1)=\lambda^l(t_2)=0$. As for the general case, in the following
section we will show that admissible variations always exist in vakonomic
mechanics.

Condition (\ref{des}) is very useful to study variation fields in vakonomic
mechanics, as we show in the following example.


\bigskip
\noindent
{\bf Example} \
Let $Q=\R^2$ be the configuration space, with coordinates $(x,y)$, and
consider a Lagrangian function and a constraint both depending only on the
velocities, i.e., $L=L(\dot{x},\dot{y})$ and $\phi = \phi(\dot{x},\dot{y})$.

From our assumptions on the constraints we can write locally $\phi =
\dot{y}-f(\dot{x})\equiv 0$. In this case, using theorem \ref{importante},
it is a simple calculus to show that the equations of motion of the
vakonomic problem are $\ddot{x}=0$, that is $x(t)=a t+b$ and $y(t)=f(a)
t+c$. The parameters $a$, $b$ and $c$ are obtained from the boundary
conditions $x(t_1)$, $y(t_1)$, $x(t_2)$ and $y(t_2)$. If
$\gamma(t)=(x(t),y(t))$ is a straight line satisfying the boundary
conditions, using (\ref{des}), the reader can check that there exist
admissible variation fields, and they are vector fields along $\gamma(t)$ of
the form $\lambda (t) \bfw$, where $\lambda(t_1)=\lambda(t_2)=0$ and
$\bfw(x,y)=(x,y;1,f'(a))$.

For example, if $\phi(\dot{x},\dot{y})\equiv \dot{y}-\sqrt{1+\dot{x}^2}=0$
and $x(t)$ is a linear function of time, there are not strongly admissible
variations (see \cite{Ar}).
But there exist admissible variation fields, so there are weakly admissible
variations. In fact, the weakly admissible variations
$\Gamma(\eps,t)=(x(\eps,t),y(\eps,t))$ have the form
$x(\eps,t)=x(t)+\lambda(t)\eps + o(\eps)$,
$y(\eps,t)=y(t)+(a/\sqrt{1+a^2})\lambda (t)\eps+o(\eps)$, where
$\lambda(t_1)=\lambda(t_2)=0$.

\subsection{The equations of motion of vakonomic mechanics}

As we have shown, a critical path of the action with constraints is an
admissible path $\gamma$ such that $\ds \int_I \left\langle
\EL_L(\ddot\gamma(t)) , \bfw(t) \right\rangle \dif t$ vanishes for each {\it
admissible} variation field~$\bfw$ (theo\-rem \ref{ELgen}). To obtain the
corresponding Euler-Lagrange equation, we first establish the following
proposition.
\begin{proposition}\label{importante0}
Given a variational problem $(Q,L,C,{\cal C},\cal{W})$, where $\cal{W}$ are
the va\-ria\-tion fields satisfying definition \ref{varvak}, let $\gamma$ be
an admissible path. Then, for any family of functions $\mu_i(t)$,
$i=1,\dots,m$, the first-order variations of the $\int_{\dot\gamma}L\,\dif
t$ and $\int_{\dot\gamma} (L + \sum_{i=1}^{m} \mu_i \phi^i) \,\dif t$ with
respect to an admissible variation field $\bfw \in\cal{W}$ coincide.
\end{proposition}

{\bf Proof:} In principle, since the variations may not be strongly
admissible, it is not clear that the variations of both actions yield the
same result. However, using theorem \ref{ELgen} and definition \ref{EL}, the
difference of the first-order variations of the actions is
$$
\int_I
\left\langle \EL_{\sum_{i=1}^{m} \mu_i \phi^i} \circ \ddot\gamma, \bfw
\right\rangle \dif t
=
\sum_{i=1}^{m} \int_I \mu_i
\left\langle \dif \phi^i \circ \dot\gamma, \invol \circ \dot\bfw
\right\rangle\dif t
-
\sum_{i=1}^{m} \int_I
\Dif \left\langle \mu_i (\FD \phi^i \circ \dot\gamma), \bfw
\right\rangle\dif t\; ,
$$
and both terms vanish whenever $\bfw$ is an admissible variation field.
Therefore, the variations of the two actions coincide when $\bfw$ is an
admissible variation field.
\hfill
$\Box$
\begin{theorem}\label{importante}
Given a variational problem $(Q,L,C,{\cal C},\cal{W})$, where $\cal{W}$ are
the va\-ria\-tion fields satisfying definition \ref{varvak}, let $\gamma$ be
an admissible path. Then $\gamma$ is critical if and only if there exist
functions $\mu_j \colon I\to\R$, $j=1,\dots m$, such that
\begin{equation}\label{VAK}
\EL_{L + \sum_{i=1}^{m} \mu_i \phi^i} \circ \ddot\gamma = 0 \; .
\end{equation}
This is the equation of motion of vakonomic mechanics.
\end{theorem}
{\bf Proof:} If equation (\ref{VAK}) holds then, for each admissible
variation field $\bfw$, $\int_I\langle \EL_{L + \sum_{i=1}^{m} \mu_i \phi^i}
\circ \ddot\gamma , \bfw \rangle\,\dif t = 0$, which, according to
Proposition (\ref{importante0}), is equivalent to $\int_I\langle \EL_{L}
\circ \ddot\gamma , \bfw \rangle\,\dif t = 0$. This shows that $\gamma$ is a
critical path.
\medskip
So it remains to prove the converse: that equation (\ref{VAK}) is a
necessary condition for the criticity of an admissible path $\gamma$.
\medskip
First, notice that the $\mu_i$ can be chosen such that
\beq
\langle
\EL_{L + \sum_{i=1}^{m} \mu_i \phi^i} \circ \ddot\gamma , \bfw_j
\rangle
= 0
\label{mu-part}
\eeq
for $j=1,\dots ,m$.
Indeed, by lemma \ref{mu-f} this equation can be written as
$$
\langle \EL_L \circ \ddot\gamma , \bfw_j \rangle +
\sum_{i=1}^{m}
\mu_i \,\langle \EL_{\phi^i} \circ \ddot\gamma , \bfw_j \rangle -
\sum_{i=1}^{m}
\Dif \mu_i \,\langle \FD \phi^i \circ \dot\gamma , \bfw_j \rangle
= 0
$$
for each $j=1,\dots ,m$. From definition \ref{EL} and the choice of $\bfw_j$
we have
\[
\langle \EL_{\phi^i} \circ \ddot\gamma , \bfw_j \rangle =
\langle \dif \phi^i \circ \dot\gamma , \invol \circ \dot\bfw_j \rangle
-
\Dif \langle \FD \phi^i \circ \dot\gamma , \bfw_j \rangle = \langle \dif
\phi^i \circ \dot\gamma , \invol \circ \dot\bfw_j \rangle \; ,
\]
for $j=1,\dots,m$. That is, we have
\begin{equation}\label{hummm}
(\Dif \mu_i) \,A^i_{\,j} - \mu_i \,B^i_{\,j} - D_j = 0\; ,
\end{equation}
where $A$ and $B$ are the matrices we have used previously (\ref{humm}). So
again we have a linear differential equation that determines the functions
$\mu_i$ (up to initial conditions). From now on we assume that $A$ is the
identity matrix; this can be easily done through a linear change of the
basis $(\bfw_i)$.

\medskip

If we apply the variational principle for the modified Lagrangian $L +
\sum_{i=1}^{m} \mu_i \phi^i$, we have
$$
\sum_{k=1}^{n}
\int_{I} \langle
\EL_{L + \sum_{i=1}^{m} \mu_i \phi^i} \circ \ddot\gamma, \bfw_k
\rangle \lambda^k\,\dif t = 0
$$
for each set of functions $\lambda^k$
yielding an admissible variation field.

If we choose the functions $\mu_i$ satisfying (\ref{mu-part}),
then the sum is only from $m+1$ to~$n$:
\begin{equation}\label{otramas}
\sum_{j=m+1}^{n}
\int_{I} \langle
\EL_{L + \sum_{i=1}^{m} \mu_i \phi^i} \circ \ddot\gamma, \bfw_j
\rangle \lambda^j\,\dif t = 0\;.
\end{equation}
This must hold for any choice of the functions $\lambda^{m+1}, \ldots
,\lambda^{n}$ giving an admissible variation field. However, as we have
shown in the preceding section, the functions
$\lambda^{m+1},\ldots,\lambda^{n}$ are not arbitrary in general, due to the
final conditions $\lambda^1(t_2)=\cdots=\lambda^m(t_2)=0$.
Let $(\bar{\mu}_i)_{1\leq i\leq m}$ be the particular solution of
(\ref{hummm}) satisfying $\bar{\mu}_i(t_2)=0$, and let $(\bar{\nu}^j_{\,i})$
be the transpose of the fundamental matrix of the homogeneous system of
(\ref{hummm}) satisfying the initial condition
$\bar{\nu}^j_{\,i}(t_2)=\delta^j_{\,i}$. Notice that $\bar{\nu}=(\nu)^{-1}$.
(In general, if $\nu$ is a fundamental matrix of $\dot{x}=A\cdot x$, then
$(\nu^t)^{-1}$ is a fundamental matrix of $\dot{x}=-A^t\cdot x$). Then the
general solution of (\ref{hummm}) is
$\mu_i=\bar{\mu}_i+\sum_{j=1}^{m}\rho_j\bar{\nu}^j_{\,i}$, where $\rho_j$
are arbitrary constants. Suppose for a while that there exist admissible
variation fields $\bfw = \sum_{k=1}^n \lambda^k\bfw_k$ with
\begin{equation}\label{fiinal}
\lambda^l=\langle \EL_{L + \sum_{i=1}^{m} (\bar{\mu}_i
+\sum_{j=1}^{m}\rho_j\bar{\nu}^j_{\,i})\phi^i} \circ \ddot\gamma, \bfw_l
\rangle\; ,
\end{equation}
for $l=m+1\dots,n$. Then (\ref{otramas}) is a vanishing sum of integrals of
squares, which, combined with (\ref{mu-part}), yields the equation of motion
(\ref{VAK}).
\medskip
It remains to prove that the choice of such $\lambda^l$, $l=m+1,\ldots,n$,
gives an admissible variation field. From (\ref{uultima}) and
$\bar{\nu}=(\nu)^{-1}$, the variations defined by (\ref{fiinal}) are
admissible if and only if
\begin{equation}\label{ajj}
\int^{t_2}_{t_1} \bar{\nu}(s)\, C(s)\,\, [\langle \EL_{L + \sum_{i=1}^{m}
(\bar{\mu}_i +\sum_{j=1}^{m}\rho_j\bar{\nu}^j_i)\phi^i} \circ \ddot\gamma,
\bfw_{(2)}
\rangle]^t\dif s = 0\; ,
\end{equation}
where $\bfw_{(2)}$ denotes the last $n-m$ vector fields. Now, from lemma
\ref{mu-f} and definition \ref{EL}, we have
\begin{equation}\label{ujj}
\langle \EL_{\sum_{j=1}^m\bar{\nu}^i_{\,j}\phi^j}\circ\ddot{\gamma}, \bfw_l
\rangle =
\sum_{j=1}^m\bar{\nu}^i_{\,j}\langle\EL_{\phi^j}\circ\ddot{\gamma},\bfw_l
\rangle -
\sum_{j=1}^m(\hbox{D}\bar{\nu}^i_{\,j})\langle\FD\phi^j\circ\dot{\gamma},\bf
w_l\rangle =
\sum_{j=1}^m\bar{\nu}^i_{\,j}\langle\dif\phi^j\circ\gamma,\invol\circ\bfw_l\
rangle\; .
\end{equation}
Using that
$C^i_{\,l}=\langle\dif\phi^i\circ\dot{\gamma},\invol\circ{\bf\dot{w}}_l
\rangle$ and combining (\ref{ajj}) and (\ref{ujj}), we obtain the linear
system
for the $\rho_j$
\[
\sum_{h=1}^m\rho_h\int_{t_1}^{t_2}\sum_{l=m+1}^n
\langle\EL_{\sum_{j=1}^m\bar{\nu}^i_{\,j}\phi^j}\circ\ddot{\gamma}, \bfw_l
\rangle\,\,
\langle \EL_{\sum_{k=1}^m\bar{\nu}^h_{\,k}\phi^k}\circ\ddot{\gamma}, \bfw_l
\rangle =
\]
\begin{equation}\label{sistemita}
= - \int_{t_1}^{t_2}\sum_{l=m+1}^n
\langle\EL_{\sum_{j=1}^m\bar{\nu}^i_{\,j}\phi^j}\circ\ddot{\gamma}, \bfw_l
\rangle\,\,
\langle \EL_{L+\sum_{k=1}^m\bar{\mu}_k\phi^k}\circ\ddot{\gamma}, \bfw_l
\rangle; .
\end{equation}
If this system has any solution, then we can find values for $\rho_j$,
$j=1,\dots,m$, such that the functions $\lambda^l$ defined in (\ref{fiinal})
give rise to admissible variations.
Now, we prove that this system has always solution. Consider the pre-Hilbert
space of continuous vector-valued functions ${\bf f}:[t_1,t_2]\to\R^{n-m}$
with the usual scalar product $\langle {\bf f},{\bf g}\rangle
=\sum_{l=m+1}^n\int_{t_1}^{t_2} f_l\cdot g_l$. Let $V$ be the
finite--dimensional subspace spanned by the $m$ vectors
\[{\bf e}_i =
\left(\langle\EL_{\sum_{j=1}^m\bar{\nu}^i_{\,j}\phi^j}\circ\ddot{\gamma},
\bfw_{m+1} \rangle, \ldots,
\langle\EL_{\sum_{j=1}^m\bar{\nu}^i_{\,j}\phi^j}\circ\ddot{\gamma}, \bfw_n
\rangle\right)\; ,
\]
$i=1,\ldots,m$. Then we can write the system (\ref{sistemita}) as
\[
\sum_{h=1}^m\rho_h\langle{\bf e}_i,{\bf e}_h\rangle = \langle{\bf e}_i,{\bf
v}\rangle\; ,
\]
where ${\bf v}=-(\langle
\EL_{L+\sum_{k=1}^m\bar{\mu}_k\phi^k}\circ\ddot{\gamma}, \bfw_{m+1}
\rangle,\ldots,\langle
\EL_{L+\sum_{k=1}^m\bar{\mu}_k\phi^k}\circ\ddot{\gamma}, \bfw_{n} \rangle)$.

The solutions of this system are any constants $\rho_h$ such that
$\sum_{h=1}^m \rho_h\cdot{\bf e}_h$ is the orthogonal projection of ${\bf
v}$ onto $V$. This is well defined, since $V$ is finite--dimensional.
(Notice that the $\rho_h$ may not be unique, since the ${\bf e}_h$ are not
necessarily independent). This completes the proof.
\hfill $\Box$

{\bf Remark:} Notice that, using lemma \ref{mu-f}, the equation of motion
may also be written as
\beq
\EL_L \circ \ddot\gamma =
\sum_{i=1}^{m} \left( (\Dif \mu_i)\, \FD \phi^i \circ \dot\gamma -
\mu_i\, \EL_{\phi^i} \circ \ddot\gamma \right)\; .
\label{VAK'}
\eeq
\medskip
{\bf Remark:} In the proofs of the equation of motion of vakonomic mechanics
that one can find in the literature, it is usually assumed that the
functions $\lambda^l$, $l=m+1,\ldots,n$, giving the admissible variations
are free. Then, the equation of motion is obtained as a straight consequence
of (\ref{otramas}). However, in general, these functions are not absolutely
free.

\section{The variations and equations of motion of nonholonomic mechanics}

In this section we are going to show that nonholonomic mechanics may be
understood as a variational problem.

Our initial setting is also the submanifold $C \subset \Tan Q$, which may be
locally defined by the vanishing of the constraints $\phi^i$. An {\it
admissible path}\/ is still a path $\gamma \colon I \to Q$ such that
$\dot\gamma$ takes its values in $C$. Let us define which are the admissible
variation fields in nonholonomic mechanics.

\begin{definition}\label{varnohol}
A variation field $\bfw$ of an admissible path $\gamma$
is called an admissible variation field (in nonholonomic mechanics)
if it is a section of the subbundle $\LD_\gamma^C \subset \gamma^* \Tan Q$.
That is,
\[
\vl(\dot\gamma(t),\bfw(t)) \in \Tan_{\dot\gamma(t)}(C)\; .
\]
\end{definition}

Using the constraints, equivalent statements are
\begin{equation}
\langle \dif \phi \circ \dot\gamma , \vl(\dot\gamma,\bfw) \rangle = 0\; ,
\end{equation}
or
\begin{equation}\label{varnohol1}
\langle \FD \phi \circ \dot\gamma , \bfw \rangle = 0\; ,
\end{equation}
for each constraint $\phi$.

Notice the key difference with respect to vakonomic mechanics: now the
admissibility is a $C^1(I)$-linear condition on $\bfw$. This linearity makes
things easier. Next, we obtain the equation of motion of nonholonomic
mechanics.

\begin{theorem}
Given a variational problem $(Q,L,C,{\cal C},\cal{W})$, where ${\cal W}$ are
the admissible variation fields satisfying definition \ref{varnohol}, an
admissible path $\gamma$ is critical if and only if there exist functions
$\mu_j$, $j=1,\dots m$, such that
\begin{equation}
\label{nonhol}
\EL_L \circ \ddot\gamma =
\sum_{i=1}^{m} \mu_i \, \FD \phi^i \circ \dot\gamma .
\end{equation}
\end{theorem}

{\bf Proof:}
A {\it critical path}\/ for nonholonomic mechanics
is an admissible path $\gamma$ such that
the first-order variation of the action,
$\ds
\int_I \left\langle \EL_L(\ddot\gamma(t)) , \bfw(t) \right\rangle \dif t ,
$
vanishes for each admissible variation field $\bfw$.
By equation (\ref{varnohol1}), this means that
$\EL_L(\ddot\gamma(t))$
is a linear combination of the
$\FD \phi^i \circ \dot\gamma$. Thus, the result follows.
\hfill $\Box$

\bigskip

{\bf Remark:} It is obvious that there always exist admissible variation
fields in nonholonomic mechanics. For example, if we calculate the
admissible variation fields of the example in section~3, we will find that,
in this case, they coincide with the admissible variation fields of
vakonomic mechanics. But this will not be true in general if the constraints
are not integrable.

\section{The case of integrable constraints}

Let us consider the problem of holonomic constraints in the usual sense.

\begin{definition} Given a differentiable manifold $Q$ and a Lagrangian
function $L \colon \Tan Q\longrightarrow\R$, a {\em holonomic problem} is a
variational problem where
\begin{itemize}
\item The constraint submanifold is given by a submanifold $P\subset Q$,
thus $C=\Tan P$.
\item Admissible paths are paths $\gamma \colon I\longrightarrow P\subset
Q$.
\item Variation fields $\bfw$ along $\gamma$ are admissible if they are
tangent to $P$.
\end{itemize}
\end{definition}

Notice that, from any admissible variation field along an admissible path
$\gamma$, one may construct a variation $\Gamma$ contained in $P$.
Therefore, it is clear that the problem with holonomic constraints is
equivalent to the unconstrained variational problem defined on $P$ by taking
the restriction of the Lagrangian $L$ to $\Tan P\subset \Tan Q$.

Now, consider the cases of both vakonomic and nonholonomic mechanics when
the constraints are defined by an integrable subbundle $C\subset \Tan Q$. In
this situation, we have the following equivalence.

\begin{theorem} If the constraint submanifold $C$ is an integrable subbundle
of $\Tan Q$, then both vakonomic and nonholonomic mechanics coincide, and
they are equi\-va\-lent to a holonomic constrained problem on each integral
submanifold of $C$.
\end{theorem}

{\bf Proof:} First of all, notice that, in both cases, a path $\gamma \colon
I\longrightarrow Q$ is admissible (i.e., $\dot{\gamma}$ is in $C$) if and
only if it is contained in an integral submanifold $P\subset Q$ of~$C$.
Recall that $\Tan_q(P)=C_q$, for each point in $P$.

\medskip

Let $\bfw$ be a variation field of $\gamma$. We know that $\bfw$ is
admissible in the nonholonomic framework if $\vl(\dot\gamma(t),\bfw(t)) \in
\Tan_{\dot\gamma(t)}(C)$ (see definition \ref{varnohol}). Since the vertical
lift restricts naturally to subbundles, this is equivalent to say that
$\vl(\dot\gamma(t),\bfw(t)) \in \hbox{V}_{\dot{\gamma}(t)} C\subset
\hbox{V}_{\dot\gamma(t)}(\Tan Q)$, that is, $\bfw (t)\in C_{\gamma(t)}$ or,
what is the same, $\bfw (t)\in \Tan_{\gamma(t)} P$. And this last condition
says that $\bfw$ is admissible for the holonomic problem. Therefore, we have
proved that a nonholonomic problem with integrable constraints is equivalent
to a holonomic problem on each integral submanifold of~$C$.

\medskip

Now, we show the equivalence with the vakonomic problem. Since $C$ is an
integrable subbundle of $\Tan Q$, it is known that the integral submanifolds
of $C$ can be locally described as $\psi=\hbox{constant}$, for some
independent functions $\psi$ on~$Q$. This implies that the constraint
submanifold $C$ can be locally described by $\phi=\hbox{\~{d}}\psi=0$. Here,
$\hbox{\~{d}}\psi \colon \Tan Q\longrightarrow\R$ is the differential
d$\psi$ of $\psi \colon Q\longrightarrow\R$ considered as a function on the
tangent bundle.

Let $\bfw$ be a variation field of $\gamma$, in the sense of vakonomic
mechanics. Then, for any $\phi=\hbox{\~{d}}\psi$, we have
$$0=\langle \hbox{d}\phi\circ\dot{\gamma},\invol\circ{\bf \dot{w}}\rangle =
\hbox{\~{d}\~{d}}\psi\circ\invol\circ{\bf\dot{w}}\; .$$
Using the property $\hbox{\~{d}\~{d}}\psi =
\hbox{\~{d}\~{d}}\psi\circ\invol$, we obtain
$$0=\hbox{\~{d}\~{d}}\psi\circ{\bf\dot{w}} =
\langle\hbox{d\~{d}}\psi\circ{\bf w},{\bf\dot{w}}\rangle =
\langle\hbox{d}\phi\circ{\bf w},{\bf\dot{w}}\rangle = \Dif (\phi\circ\bfw) =
\Dif\langle\hbox{d}\psi\circ\gamma,\bfw\rangle\; .$$
Thus, $\bfw$ is an admissible variation field in vakonomic mechanics if and
only if $\langle\hbox{d}\psi\circ\gamma,\bfw\rangle$ is constant. Since
$\bfw (t_1)=0$, this constant is zero,
$\langle\hbox{d}\psi\circ\gamma,\bfw\rangle=0$, which means that $\bfw$ is
tangent to $P$. Therefore, $\bfw$ is admissible for the vakonomic problem if
and only if it is admissible for the holonomic problem on $P$.
\hfill $\Box$

\section{Optimal control and vakonomic mechanics}

A problem of optimal control may be given by the following data: a
configuration space $B$ describing the state variables, a fibre bundle
$\pi\colon M\longrightarrow B$ whose fibres describe the control variables,
a vector field $Y$ along the projection of the bundle, $Y\colon
M\longrightarrow\Tan B$, and a ``Lagrangian function" $L\colon
M\longrightarrow\R$. For a path $\gamma\colon I\to M$ where $\pi\circ\gamma$
(not $\gamma$!) has fixed end-points, the problem is to find an extremum of
the action
\[
\int_\gamma L(\gamma(t))\,\dif t
\]
when $\gamma$ satisfies the differential equation
\begin{equation}\label{difcon}
(\pi\circ\gamma)^{\textstyle.} = Y\circ\gamma
\end{equation}
that rules the evolution of the state variables.

It is easy to show that this is indeed a vakonomic problem on the
manifold~$M$, in which the Lagrangian $L$ is very singular, since it does
not depend on the velocities. The constraint submanifold $C\subset \Tan M$,
given by the differential equation above, is
\begin{equation}
C=\{ w_u\in \Tan M\mid\Tan\pi (w_u)=Y(u)\}\; .
\end{equation}
In this way, a path $\gamma$ is admissible if and only if it is a solution
of the differential equation (\ref{difcon}) or, equivalently, if it takes
values in the affine subbundle $C$ of $\Tan M$. On the other hand, from the
special characteristics of optimal control problems, we can relax the
boundary conditions as we have done. Observe that theorems \ref{ELgen} and
\ref{importante} remain true since $L$ does not depend on the velocities and
the structure of the constraints (\ref{difcon}) (they do not depend on the
derivatives on the control variables). In coordinates, if $x^i$, $i=1,\dots
m$, are local coordinates in $B$ and $(x^i,u^\alpha)$, $i=1,\dots m$,
$\alpha=1,\dots,n-m$, are local coordinates in $M$, the action is given by
\[
\int_{t_1}^{t_2} L(x^i,u^\alpha)\,\dif t
\]
and the constraints are given by a set of first order differential equations
\[
\dot{x}^i=Y^i(x,u)~,~~i=1,\dots,m,
\]
with boundary conditions $x^i_1=x^i(t_1)$ to $x^i_2=x^i(t_2)$ (there are no
boundary conditions on the control variables). Notice that for this
vakonomic problem the constraints are very particular: they express the
velocities of the state variables in terms of the state and control
variables.

\medskip

Let us identify which variation fields $\bfw$ are admissible. Notice that
frames for the bundles $\LD_\gamma^C \subset \gamma^* \Tan(M)$ are provided
with $(\bfw_\alpha)$ and $(\bfw_i,\bfw_\alpha)$, where for instance
$$
\bfw_\alpha = \derpar{}{u^\alpha} \circ \gamma , \quad
\bfw_i = \derpar{}{x^i} \circ \gamma ,
$$
if we have coordinates $(x^i,u^\alpha)$ on~$M$.
Writing $\bfw = \lambda^i \bfw_i + \lambda^\alpha \bfw_\alpha$,
the differential equation (\ref{des}) turns out to be
$$
\Dif \lambda^i - \derpar{y^i}{x^j} \lambda^j = \derpar{y^i}{u^\alpha}
\lambda^\alpha .
$$

\medskip

Optimal control theory admits several geometric formulations and expressions
of the equation of motion. First, we can give a {\it lagrangian
description}\/ on the configuration manifold $\Tan^*B \times_B M$. We only
need to define a lagrangian on its tangent space:
$$
{\cal L}(x,p,u,\dot{x},\dot p,\dot u) = L(x,u) + \langle p, \dot x - Y(x,u)
\rangle
$$
where we write $(x;p,u)$ for the variables of $\Tan^*B \times_B M$---recall
that $\Tan(\Tan^*B \times_B M) = \Tan(\Tan^*B) \times_{\Tan B} \Tan M$; its
elements are pairs of tangent vectors $(\dot{x},\dot p,\dot u)$ projecting
to the same tangent vector $\dot x$. For a path $\eta$ on $\Tan^*B \times_B
M$ the Euler-Lagrange equation $\EL_{\cal L} \circ \ddot\eta$ is readily
seen to be equivalent to the vakonomic equation (\ref{VAK}).

\medskip

Let us recall that, given a $l$-dimensional differentiable manifold $Q$ with
local coordinates $(q^A)$, there is a canonical tensor field on $\Tan Q$,
the vertical endomorphism $S$ which is a rank-$l$ (1,1) tensor field on
$\Tan Q$ such that $\Ker S = \hbox{Im}S$ and whose Nijenhuis tensor $N_S$
vanishes. In natural coordinates $(q^A,\dot{q}^A)$, the local expression of
$S$ is given by $S=\dif q^A\otimes\frac{\partial}{\partial\dot{q}^A}$. Also,
we have the Liouville vector
field $\Delta$ (the infinitesimal generator of the dilations along the
fibres on $\Tan Q$), whose local expression is $\Delta=\dot{q}^A\partial
/\partial\dot{q}^A$. A vector field $X$ in $\Tan Q$ is called a second order
differential equation (SODE) if $S(X)=\Delta$. Now, if ${\cal L} \colon \Tan
Q\longrightarrow\R$ is a Lagrangian function, we can construct the Cartan
1-form associated with ${\cal L}$, given by $\theta_{\cal L}=S^*\circ\dif
{\cal L}$, the Cartan 2-form $\omega_{\cal L}=-\dif\theta_{\cal L}$ and the
energy function $E_{\cal L}=\Delta ({\cal L})-{\cal L}$. Then the paths
$\eta$ solution of the Euler-Lagrange equations are the integral curves of a
second order differential equation $X$ in $\Tan Q$ satisfying the dynamical
equation $i_X\omega_{\cal L}=\dif E_{\cal L}$.

\medskip

Taking $Q=\Tan^*B\times_B M$ and ${\cal L}= L(x,u) + \langle p, \dot x -
Y(x,u)\rangle$, we obtain a geometrical expression of the equation of motion
of optimal control theory,
\[
i_X\omega_{\cal L}=\dif E_{\cal L}\; ,
\]
where $E_{\cal L}=\langle p, Y(x,u)\rangle - L(x,u)$ and $\omega_{\cal
L}=-\dif\theta_{\cal L} = -\dif(p_i \dif x^i) = \dif x^i\wedge\dif p_i$.

\medskip
Associated to this lagrangian description we can consider the hamiltonian
formalism of~${\cal L}$. This means to consider the manifold $\Tan^*(\Tan^*B
\times_B M)$ with its canonical symplectic structure, the Legendre's
transformation of~${\cal L}$, $\FD{\cal L} \colon \Tan(\Tan^*B \times_B M)
\to \Tan^*(\Tan^*B \times_B M)$, and to push forward through it the energy
function to a hamiltonian function ${\cal H} \colon \Tan^*(\Tan^*B \times_B
M) \to \R$.

\medskip

However, the most interesting geometric description of optimal control
theory is a {\it presymplectic description}\/ which can be constructed on
the manifold $\Tan^*B \times_B M$.
Here we consider the 2-form $\omega$ obtained by pull-back through
$\Tan^*B \times_B M \to \Tan^*B$ of the canonical 2-form of the last
manifold. In local coordinates, $\omega=\dif q \wedge \dif p$. Taking the
hamiltonian function defined by
$$
H(x,u,p) = \langle p,Y(x,u) \rangle - L(x,u)\; ,
$$
if $\eta$ is a path on $\Tan^*B \times_B M$, the presymplectic equation
$$
i_{\dot\eta} \omega = \dif H \circ \eta
$$
is equivalent to the equation of motion of vakonomic mechanics,
in the sense that there is a natural bijection between
both sets of solutions. To show this is enough to write the local
expressions.

\medskip
{\bf Remark:} In optimal control theory the hamiltonian function is usually
written as $H(x,u,p) = \langle p,Y(x,u) \rangle - \mu_0 L(x,u)$, where
$\mu_0=0,1$. When $\mu_0=0$ we recover the so called abnormal solutions (see
\cite{LiuS}). However, in the vakonomic approach, there are not abnormal
solutions. The key issue is that we work with admissible variation fields,
not admissible variation curves.

\section{Constraints defined by a distribution}

In this section we present a geometric framework for constrained systems
when the cons\-traint submanifold $C$ is a distribution
(or vector subbundle) of the tangent ma\-ni\-fold $\Tan Q$.
In local coordinates, this means that the constraints are linear functions
on the velocities.
The subbundle $C\subset\Tan Q$ can be described in terms of its annihilator,
$C^0\subset\Tan^* Q$.
If this is locally described in terms of 1-forms,
$\alpha^i=\alpha^i_a(q)\dif q^a$
($i=1,\dots,m$, where $m$ is the codimension of $C$ and $(q^a)$ are local
coordinates of $Q$), then $C$ is locally described in terms of the
constraints
\[
\phi^i(v_q) = \langle \alpha^i(q),v_q \rangle = 0 , \quad i=1,\dots, m\; .
\]

\smallskip
Let us consider the vector bundle $\Tan Q\oplus C^0$, in which we will set
up the dynamics.

On the one hand, given the Lagrangian function $L$ on the tangent bundle
$\Tan Q$, let $\theta_L=S^*\circ\dif L$ be the Lagrange 1-form on $\Tan Q$.
Its pull-back along the projection
$\pi_1\colon\Tan Q\oplus C^0\longrightarrow\Tan Q$
yields the 1-form
$$
\theta_1=\pi_1^*\theta_L
$$
on $\Tan Q\oplus C^0$.
Also, using the Liouville vector field $\Delta$ on $\Tan Q$, the energy
function associated with $L$ in $\Tan Q\oplus C^0$ is
$$
E=\pi_1^*(\Delta (L)-L) .
$$

On the other hand, let $\theta_Q$ be the canonical 1-form defined on the
cotangent bundle $\Tan^*Q$.
If $j_0\colon C^0\longrightarrow \Tan^*Q$ denotes the canonical inclusion
and $\pi_2 \colon\Tan Q\oplus C^0\longrightarrow C^0$ is the projection onto
the second factor, then we can take the pull-back of these mappings to
construct a 1-form $\theta_2$ on $\Tan Q\oplus C^0$ as
$$
\theta_2=(j_0\circ\pi_2)^*\theta_Q .
$$

Using the 1-forms $\theta_1$ and $\theta_2$, we have a presymplectic form
$$
\Omega=-\dif(\theta_1+\theta_2) .
$$
By means of the energy function $E$, we obtain a presymplectic dynamics on
the extended phase space $\Tan Q\oplus C^0$ which is equivalent to vakonomic
mechanics:

\begin{theorem}
\label{vaklineal}
Let $L \colon \Tan Q \to \R$ be a Lagrangian,
and $C \subset \Tan Q$ a vector subbundle.
For a path $\xi$ in the manifold $\Tan Q \oplus C^0$,
consider the differential equation
\beq
i_{\dot\xi} \Omega = \dif E \circ \xi .
\label{vaksum}
\eeq
This equation is equivalent to the equation of motion
of vakonomic mechanics in the following sense:
\begin{itemize}
\item
If $\xi = (\xi_1,\xi_2)$ is a solution of (\ref{vaksum})
and $\xi_1$ is the lift of a path in~$Q$,
$\xi_1 = \dot\gamma$,
then $\gamma$ is an admissible path
($\dot\gamma$ is in~$C$) and is a solution of
the equation of motion of vakonomic mechanics (\ref{VAK}).
\item
Conversely, given an admissible path $\gamma$ which is a solution of
(\ref{VAK}),
together with the multipliers $\mu^i$,
then the path
$\xi(t) = (\dot\gamma(t), \sum \mu_i(t) \dif \phi^i(\dot\gamma(t)))$
is a solution of equation (\ref{vaksum}).
\end{itemize}
If the Lagrangian is regular then equation (\ref{vaksum}) already implies
that $\xi_1$ is the lift of a path in~$Q$.
\end{theorem}

{\bf Proof:}
It is enough to check the equivalence in local coordinates.
We take $(q^a,v^a,\lambda_i)$, $a=1,\dots,n$, $i=1,\dots,m$,
as local coordinates in $\Tan Q \oplus C^0$
(we represent an element of $C^0_q$ as $\sum \lambda_i \alpha^i(q)$).
Then we have
\begin{eqnarray*}
\theta_1 &=& \frac{\partial L}{\partial v^a} \dif q^a ,
\\
\theta_2 &=& \lambda_i \alpha^i_a(q) \dif q^a ,
\\
\Omega &=&
\left( \frac{\partial^2 L}{\partial v^a\partial q^b} +
\lambda_i \frac{\partial \alpha^i_a}{\partial q^b} \right)
\dif q^a \wedge \dif q^b
+
\frac{\partial^2 L}{\partial v^a \partial v^b}
\dif q^a \wedge \dif v^b
+
\alpha^i_a \dif q^a \wedge \dif \lambda_i .
\end{eqnarray*}
Since
$E = v^a \left(\partial L/\partial v^a\right) - L$,
we also have
$$
\dif E =
\left( v^b \frac{\partial^2 L}{\partial q^a\partial v^b} -
\frac{\partial L}{\partial q^a} \right) \dif q^a +
v^b \frac{\partial^2 L}{\partial v^a \partial v^b} \dif v^a .
$$

Now let us consider the path $\xi(t) = (q^a(t),v^a(t),\lambda^i(t))$,
with velocity
$\dot\xi = (q,v,\lambda;\dot q,\dot v,\dot \lambda)$.
Then
\begin{eqnarray}
i_{\dot\xi}\Omega & = &
\left(
\dot q^b \frac{\partial^2 L}{\partial q^a \partial v^b}
+ \dot q^b \frac{\partial \alpha^i_b}{\partial q^a}\lambda_i
- \dot q^b \frac{\partial^2 L}{\partial v^a \partial q^b}
- \dot q^b \frac{\partial \alpha^i_a}{\partial q^b}\lambda_i
- \dot v^b \frac{\partial^2 L}{\partial v^a \partial v^b}
- \dot \lambda_i \alpha^i_a
\right) \dif q^a
\nonumber \\
&& \hbox{}
+ \dot q^b \frac{\partial^2 L}{\partial v^a \partial v^b} \dif v^a
+ \dot q^a \alpha^i_a \dif \lambda_i .
\nonumber
\end{eqnarray}

Therefore, equation $i_{\dot\xi}\Omega = \dif E$ is equivalent
to the three equations
\begin{equation}
\label{un}
(\dot q^b-v^b) \frac{\partial^2 L}{\partial q^a \partial v^b}
+ \dot q^b \frac{\partial \alpha^i_b}{\partial q^a} \lambda_i
- \dot q^b \frac{\partial^2 L}{\partial v^a \partial q^b}
- \dot q^b \frac{\partial \alpha^i_a}{\partial q^b} \lambda_i
- \dot v^b \frac{\partial^2 L}{\partial v^a \partial v^b}
- \dot \lambda_i \alpha^i_a
= - \frac{\partial L}{\partial q^i} ,
\end{equation}
\begin{equation}
\label{dos}
\dot q^b \frac{\partial^2 L}{\partial v^a \partial v^b}
=
v^b \frac{\partial^2 L}{\partial v^a \partial v^b} ,
\end{equation}
\begin{equation}
\label{tres}
\dot q^a \alpha^i_a = 0 .
\end{equation}

The fact that $\xi_1$ is the lift of a path $\gamma$ in~$Q$
means in coordinates that $v(t) = \dot q(t)$,
so equation (\ref{dos}) is an identity.
Notice also that if the Lagrangian is regular then the Hessian matrix
$\left(\frac{\partial^2 L}{\partial v^a \partial v^b}\right)$
is invertible, therefore in this case equation (\ref{dos}) implies that
$v(t) = \dot q(t)$,
that is to say,
$\xi_1$ is the lift of a path in~$Q$.
Then, in equation (\ref{tres}) we obtain the constraints
$\phi^i(q,\dot q) = \alpha^i_a(q) \dot{q}^a = 0$, that is,
$\gamma$ is an admissible path.
Finally, we can write equation (\ref{un}) as
\[
\ddot q^b \frac{\partial^2 L}{\partial v^a \partial v^b} +
\dot q^b \frac{\partial^2 L}{\partial v^a \partial q^b} +
\dot q^b \frac{\partial \alpha^i_a}{\partial q^b} \lambda_i +
\dot \lambda_i \alpha^i_a
=
\frac{\partial L}{\partial q^a} +
\dot q^b \frac{\partial \alpha^i_b}{\partial q^a} \lambda_i .
\]
But these are the vakonomic equations (\ref{VAK}) of the extended Lagrangian
${\cal L} = L + \mu_i \alpha^i_a v^a$,
using the natural identification between the functions $\mu_i$
and the coordinates $\lambda_i$ of the cotangent vectors.
\hfill
$\Box$

\bigskip

{\bf Remark:}
In a similar way, the vakonomic dynamics can be also defined on the manifold
$C\oplus C^0$.
Since $C\oplus C^0$ is a vector subbundle of $\Tan Q\oplus C^0$, we can
pull-back the $2$-form $\Omega$ and the energy function $E$ to define a
$2$-form $\tilde{\Omega}$ and a new function $\tilde{E}$ in $C\oplus C^0$.
The reader can check that,
then the equation of motion of vakonomic mechanics (\ref{VAK})
is equivalent to
find the paths $\xi = (\xi_1,\xi_2)$ in $C\oplus C^0$,
where $\xi_1$ is the lift of a path in~$Q$,
such that
\begin{equation}
\label{vakCC}
i_{\dot\xi} \tilde{\Omega} = \dif\tilde{E} \circ \xi .
\end{equation}
Moreover,
if the lagrangian is regular,
then $\tilde\Omega$ is a symplectic form.
Notice also that this equation, as well as equation (\ref{vaksum}),
can also be expressed in terms of vector fields.
For instance, when $\tilde\Omega$ is symplectic,
the solutions of equation (\ref{vakCC})
are the integral curves of the vector field $\tilde X$ such that
$$
i_{\tilde X} \tilde{\Omega} = \dif\tilde{E} .
$$
\medskip

In the case of nonholonomic mechanics,
a similar result can be proved,
in the same way as for theorem~\ref{vaklineal}.
Let us denote $\Omega_1 = -\dif \theta_1$.
Then we have:
\begin{theorem}
\label{nhlineal}
Let $L \colon \Tan Q \to \R$ be a Lagrangian,
and $C \subset \Tan Q$ a vector subbundle.
For a path $\xi$ in the manifold $\Tan Q \oplus C^0$,
consider the differential equation
\beq
i_{\dot\xi} \Omega_1 = \dif E \circ \xi + \theta_2 \circ \xi .
\label{nhsum}
\eeq
This equation is equivalent to the equation of motion
of nonholonomic mechanics in the following sense:
\begin{itemize}
\item
If $\xi = (\xi_1,\xi_2)$ is a solution of (\ref{nhsum})
where $\xi_1$ is the lift of a path $\gamma$ in~$Q$,
$\xi_1 = \dot\gamma$,
then $\gamma$ is a solution of
the equation of motion of nonholonomic mechanics (\ref{nonhol}).
\item
Conversely, given a path $\gamma$ which is a solution of (\ref{nonhol}),
together with the multipliers $\mu^i$,
then the path
$\xi(t) = (\dot\gamma(t), \sum \mu_i(t) \dif \phi^i(\dot\gamma(t)))$
is a solution of equation (\ref{nhsum}).
\end{itemize}
If the Lagrangian is regular then equation (\ref{nhsum}) already implies
that $\xi_1$ is the lift of a path in~$Q$.
\hfill$\Box$
\end{theorem}

\section{Conclusions}

In this paper we have presented variational calculus
(in one dimension)
in a geometric framework,
aiming to study dynamical systems with non-holonomic constraints
({\it i.e.}, constraints depending on the positions and the velocities).
We have shown that a generalised formulation of variational calculus,
in which the admissible paths and the admissible infinitesimal variations
are not necessarily related,
makes room for the study of dynamical systems subject to
non-holonomic constraints from different points of view.
This generalized variational calculus encompasses
the often-called vakonomic mechanics
(which is a strict variational problem with constraints)
and the non-holonomic mechanics
(based on d'Alembert's principle).

In the case of vakonomic mechanics, we have provided a geometric procedure
to obtain the equation of motion, choosing an appropriate set of admissible
infinitesimal variations proving that they always exist.

In the case of non-holonomic mechanics, it is far more simple than in
vakonomic mechanics to choose an appropriate set of admissible infinitesimal
variations, and the corresponding equation of motion is readily obtained.

Our formulation also provides a neat equivalence between both
vakonomic and non-holonomic mechanics when the constraints are integrable
(also called holonomic).

We have also found the geometry lying on some particular cases of
vakonomic mechanics,
namely the case of optimal control and
the case where the constraint submanifold is a vector subbundle of the
tangent bundle.

All the paper is written for the case of time-independent
lagrangian and constraints,
but the reader may check that the time-dependent case
may be dealt with
by adjunction of the time variable
in a not too involved way.

\section*{Acknowledgements}
The authors thank N. Rom\'an-Roy for useful discussions.
X.G. and M.C.M.L. acknowledge partial financial support
from CICYT TAP 97--0969--C03--01 and PB98--0920. J.M.S. acknowledge
partial financial support from CICYT projects PB98--0821 and PB98--0920.


\end{document}